\documentclass[conference]{IEEEtran}
\IEEEoverridecommandlockouts
\usepackage{cite}
\usepackage{amsmath,amssymb,amsfonts}
\usepackage{algorithmic}
\usepackage[ruled]{algorithm2e}
\usepackage{graphicx}
\usepackage{textcomp}
\usepackage{xcolor}
\def\BibTeX{{\rm B\kern-.05em{\sc i\kern-.025em b}\kern-.08em
    T\kern-.1667em\lower.7ex\hbox{E}\kern-.125emX}}
\begin{document}

\title{Delay Alignment Modulation with Hybrid Beamforming for Spatially Sparse Communications\\
\thanks{This work was supported by the National Key R\&D Program of China with Grant number 2019YFB1803400.}
}

\author{\IEEEauthorblockN{Jieni Zhang$^{*}$ and Yong Zeng$^{*\dag }$}
\IEEEauthorblockA{$^{*}$National Mobile Communications Research Laboratory, Southeast University, Nanjing 210096, China}
\IEEEauthorblockA{$^{\dag}$Purple Mountain Laboratories, Nanjing 211111, China}
\IEEEauthorblockA{jieni\_zhang@163.com, yong\_zeng@seu.edu.cn}
}

\maketitle

\begin{abstract}
For millimeter wave (mmWave) or Terahertz (THz) communications, by leveraging the high spatial resolution offered by large antenna arrays and the multi-path sparsity of mmWave/THz channels, a novel inter-symbol interference (ISI) mitigation technique called delay alignment modulation (DAM) has been recently proposed. The key ideas of DAM are \textit{delay pre-compensation} and \textit{path-based beamforming}. However, existing research on DAM is based on fully digital beamforming, where the number of radio frequency (RF) chains required is equal to the number of antennas. This leads to high hardware cost and power consumption for mmWave/THz massive multiple-input multiple-output (MIMO) communications. Thus, this paper proposes the hybrid analog/digital beamforming based DAM. The analog and digital beamforming matrices are designed to achieve performance close to DAM based on fully digital beamforming. The effectiveness of the proposed technique is verified by simulation results.
\end{abstract}


\section{Introduction}
With the increasing demand for high data rate transmission, massive multiple-input multiple-output (MIMO) with millimeter wave (mmWave) or Terahertz (THz) communication has received extensive attention during the past few years \cite{mmWave1,mmWave2}. However, for broadband communication systems, the time-dispersive channel caused by multi-path propagation results in inter-symbol interference (ISI) that impairs the system performance. To address this issue, a novel technique called delay alignment modulation (DAM) was recently proposed in \cite{DAM} based on the key ideas of \textit{delay pre-compensation} and \textit{path-based beamforming}, without relying on conventional techniques like channel equalization or multi-carrier transmission. Specifically, by leveraging the high spatial resolution of large antenna arrays \cite{spatial resolution} and the multi-path sparsity \cite{Sparsity,Sparsity-1,Sparsity-2} of mmWave/THz channels, symbol delays are purposely introduced at the transmitter aiming to compensate the delays of channel multi-paths, so that with appropriate path-based beamforming, all multi-path signal components will arrive at the receiver simultaneously and constructively. As a result, not only ISI is eliminated at the receiver, but also the channel power contributed by all multi-path components is obtained. Based on such a principle, the authors in \cite{DAM-OFDM} further proposed a more generic DAM technique to reduce the channel delay spread when complete elimination of ISI is infeasible or undesirable. This provides a new degree of freedom against time-dispersive channels, which enables more efficient single- or multi-carrier signaling transmission. Furthermore, the DAM-orthogonal frequency division multiplexing (DAM-OFDM) technique was proposed, which can save the cyclic prefix (CP) overhead or reduce the peak-to-average-power ratio (PAPR) of OFDM by reducing the channel delay spread. The application of DAM for multi-user communication or systems assisted by intelligent reflected surfaces (IRSs) was studied in \cite{DAM-mu} and \cite{DAM-IRS}, respectively. In addition, an efficient channel estimation method for DAM was proposed in \cite{DAM-ce}. Furthermore, the authors in \cite{DAM-ISAC1,DAM-ISAC2} investigated integrated sensing and communications (ISAC) based on DAM.

The aforementioned research works have demonstrated the great potential of DAM for spatially sparse channels with large antenna arrays, in terms of enhancing spectral efficiency, reducing PAPR, and simplifying receiver design. However, the aforementioned works on DAM are based on fully digital beamforming, which requires the number of radio frequency (RF) chains equal to the number of antennas \cite{Alt}. To reduce the hardware cost and power consumption, hybrid analog/digital beamforming has been extensively studied, which requires only a small number of RF chains between the digital beamformer and the analog beamformer \cite{Hybrid-ana,OMP,ULA}. 

In this paper, hybrid analog/digital beamforming based DAM was proposed, which aims to achieve performance close to that of fully digital beamforming with much lower power consumption and hardware cost. An efficient algorithm is presented to jointly optimize the baseband digital beamforming and analog beamforming for DAM, and theoretical analysis is provided for the resulting effective spectral efficiency and PAPR. Furthermore, comparisons between DAM and OFDM based on hybrid beamforming are presented. 
Finally, simulation results are provided to verify the effectiveness of hybrid beamforming based DAM and demonstrate that it achieves comparable or even better effective spectral efficiency and bit error rate (BER) than the OFDM counterpart, with much reduced PAPR. 

\section{SYSTEM MODEL}
We consider a spatially sparse wireless communication system, such as mmWave or THz system, in which a base station (BS) equipped with  ${M_\mathrm{t}} \gg 1$ antennas communicates with a single-antenna user equipment (UE). For cost-effective implementation, the BS has only ${M_{\mathrm{RF}}} < {M_\mathrm{t}}$ RF chains and hybrid analog/digital beamforming is applied. In a time-dispersive multi-path environment, the discrete-time representation of the channel impulse response can be expressed as~\cite{DAM}
\begin{equation}
	{{\mathbf{h}}^H}[n] = \sum\limits_{l = 1}^L {{\mathbf{h}}_l^H} \delta [n - {n_l}],
	\label{Channel}
\end{equation}
where $L$ denotes the number of temporal-resolvable multi-paths, ${{\mathbf{h}}_l} \in \mathbb{C}^{{M_\mathrm{t}} \times 1}$ and $n_l$ denote the channel vector and the discretized delay of the $l$th multi-path respectively. Due to the multi-path sparsity in mmWave/THz communications with large antenna arrays, it can be concluded that ${M_\mathrm{t}} \gg L$~\cite{Sparsity-1,Sparsity-2,Sparsity}. Because each temporal-resolvable multi-path in \eqref{Channel} may consist of several sub-paths that have the same delay but distinct angles of departure (AoDs), ${\mathbf{h}}_l$ can be modeled as~\cite{DAM}
\begin{equation}
{{\bf{h}}_l} = {\alpha _l}\sum\nolimits_{i = 1}^{{\mu _l}} {{v_{li}}{{\bf{a}}_\mathrm{t}}\left( {{\theta _{li}}} \right)}, 
\label{hl}
\end{equation}
where $\alpha_l$ and $\mu _l$ denote the complex path gain and the number of sub-paths of the $l$th multi-path, respectively, ${v_{li}} = \sqrt {{\varsigma _{li}}} {e^{j{\phi _{li}}}}$ denotes the complex coefficient of the $i$th sub-path of the $l$th multi-path, satisfying $\sum\nolimits_{i = 1}^{{\mu _l}} {{\varsigma _{li}} = 1} $, and $\theta_{li}$  denotes the AoD of the $i$th sub-path of the $l$th multi-path, while ${{\mathbf{a}}_\mathrm{t}}({\theta _{li}}){ \in \mathbb{C}^{{M_\mathrm{t}} \times 1}}$ denotes  the corresponding transmit array response vector. For a basic uniform linear array (ULA), the transmit array response vector ${{\mathbf{a}}_\mathrm{t}}({\theta _{li}})$ is given by
\begin{equation}
	{{\mathbf{a}}_\mathrm{t}}({\theta _{li}}) = {\left[ {1,{e^{ - j\frac{{2\pi d}}{\lambda }\sin ({\theta _{li}})}}, \cdots ,{e^{ - j\frac{{2\pi ({M_\mathrm{t}} - 1)d}}{\lambda }\sin ({\theta _{li}})}}} \right]^T},
	\label{array response}
\end{equation}
where $\lambda$  denotes the signal wavelength and $d$ denotes the inter-element spacing of the ULA.

Let ${\mathbf{x}}[n]{ \in \mathbb{C}^{{M_\mathrm{t}} \times 1}}$ denote the discrete-time equivalent of the transmitted signals by the BS. The received signal by the UE can be expressed as
\begin{equation}
	y[n] = {{\bf{h}}^H}[n] * {\bf{x}}[n] + z[n] = \sum\limits_{l = 1}^L {{\bf{h}}_l^H} {\bf{x}}\left[ {n - {n_l}} \right] + z[n],
	\label{y[n]}
\end{equation}
where $z[n] \sim {\cal C}{\cal N}(0,{\sigma ^2})$ is the additive white gaussion noise (AWGN). Let ${n_{\min }} \buildrel \Delta \over = \mathop {\min }\limits_{1 \le l \le L} {n_l}$ and ${n_{\max }} \buildrel \Delta \over = \mathop {\max }\limits_{1 \le l \le L} {n_l}$ denote the minimum and maximum delay over all the $L$ multi-paths, respectively. Thus, the channel delay spread is given by ${n_{{\rm{span}}}} = {n_{\max }} - {n_{\min }}$. In~\cite{DAM}, a novel technique called DAM was proposed to resolve the ISI issue based on \textit{delay pre-compensation} and \textit{path-based beamforming}, without relying on conventional techniques like channel equalization or multi-carrier transmission. However, the DAM technique presented there assumes the fully digital beamforming at the BS, which cannot be applied for the considered system with fewer RF chains than the number of antennas. In the following, we propose the hybrid analog/digital beamforming based DAM technique for spatially sparse communication systems.

\section{Hybrid Beamforming based DAM}
\begin{figure}[t!]
	\centerline{\includegraphics[width=8.5cm]{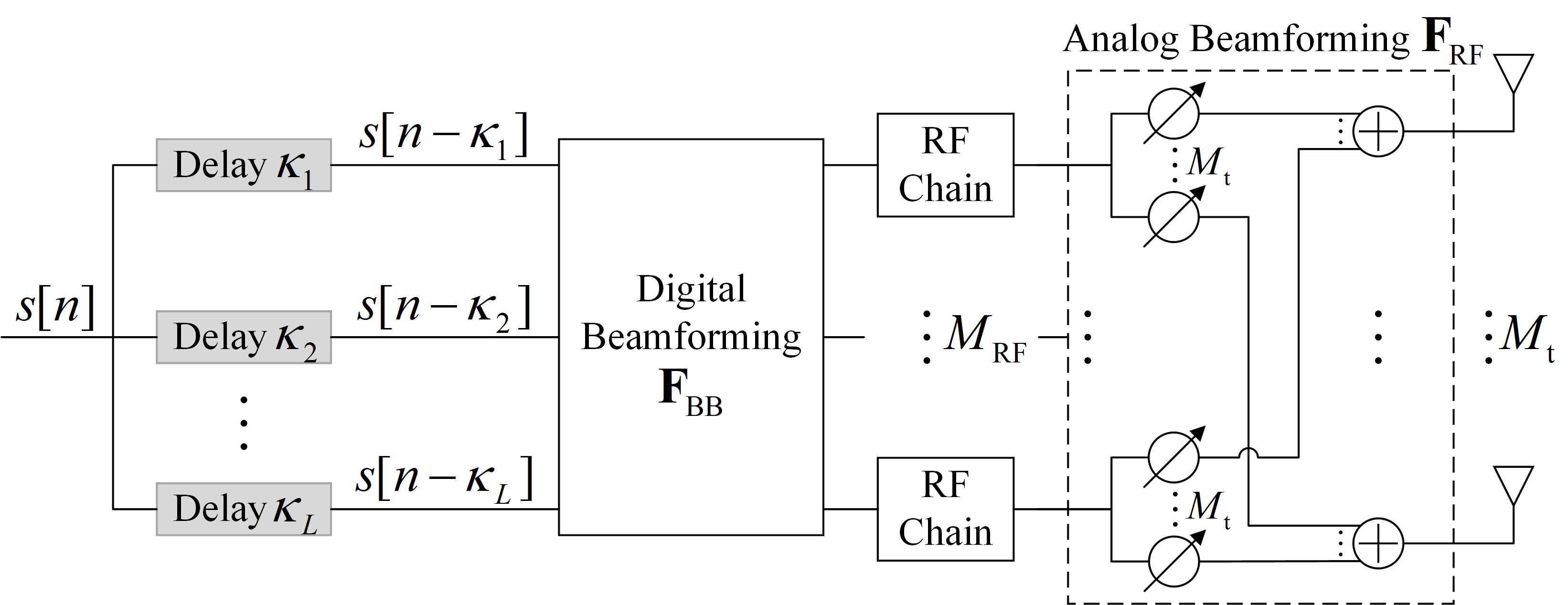}}
	\caption{Transmitter architecture for DAM based on hybrid beamforming.}
	\label{DAM}
\end{figure}

The transmitter architecture of the proposed DAM based on hybrid analog/digital beamforming is illustrated in Fig.~\ref{DAM}. The signal transmitted by the $M_\mathrm{t}$ antennas of the BS can be expressed as
\begin{equation}
	{\bf{x}}[n] = {{\bf{F}}_{\mathrm{RF}}}\sum\limits_{l = 1}^L {{{\bf{f}}_{\mathrm{BB},l}}s[n - {\kappa _l}]},
	\label{x[n]}
\end{equation}
where ${{\bf{F}}_{\mathrm{RF}}}{ \in \mathbb{C}^{{M_\mathrm{t}} \times {M_{\mathrm{RF}}}}}$ denotes the analog beamforming matrix with unit modulus for each element, i.e., $| {{{\left( {{{\bf{F}}_{\mathrm{RF}}}} \right)}_{i,j}}} | = 1,\forall i,j$, ${{\bf{F}}_{\mathrm{BB}}}{ \in \mathbb{C}^{{M_{\mathrm{RF}}} \times L}}$ denotes the digital baseband beamforming matrix that includes ${{\bf{f}}_{\mathrm{BB},l}}{ \in \mathbb{C}^{{M_{\mathrm{RF}}} \times 1}}$ for its $l$th column, $s[n]$ is the independent and identically distributed (i.i.d.) information-bearing symbols with normalized power $\mathbb{E}[ {{{| {s[ n ]} |}^2}} ] = 1$, and ${\kappa _l} \ge 0$ is the deliberately introduced delay with ${\kappa _l} \ne {\kappa _{l'}},\forall l \ne l'$, aiming to compensate for the delay $n_l$ of the $l$th channel path.
The power of the transmitted signal $\bf{x}\rm[n]$ is
\begin{equation}
	\begin{split}
		\mathbb{E}\left[ {\left\| {{\bf{x}}[n]} \right\|^2} \right] =& \sum\limits_{l = 1}^L {\mathbb{E}\left[ {\left\| {{{\bf{F}}_{\mathrm{RF}}}{{\bf{f}}_{\mathrm{BB},l}}s[n - {\kappa _l}]} \right\|^2} \right]} \\
		=& \sum\limits_{l = 1}^L {\left\| {{{\bf{F}}_{\mathrm{RF}}}{{\bf{f}}_{\mathrm{BB},l}}} \right\|^2 = \left\| {{{\bf{F}}_{\mathrm{RF}}}{{\bf{F}}_{\mathrm{BB}}}} \right\|_F^2 \le P} ,
	\end{split}
	\label{P}
\end{equation}
where $P$ denotes the available transmit power. In \eqref{P}, the first equality holds because $s[n]$ is independent for different $n$ and ${\kappa _l} \ne {\kappa _{l'}},\forall l \ne l'$. By substituting \eqref{x[n]} into \eqref{y[n]}, the received signal of DAM with hybrid analog/digital beamforming is
\begin{equation}
	\begin{split}
		y[n] = & \sum\limits_{l = 1}^L {{\bf{h}}_l^H{{\bf{F}}_{\mathrm{RF}}}{{\bf{f}}_{\mathrm{BB,}l}}s[n - {\kappa _l} - {n_l}] + } \\
		&\sum\limits_{l = 1}^L {\sum\limits_{l' \ne l}^L {{\bf{h}}_l^H{{\bf{F}}_{\mathrm{RF}}}{{\bf{f}}_{\mathrm{BB,}l'}}s[n - {\kappa _{l'}} - {n_l}]} }  + z[n].
	\end{split}
	\label{y[n] of DAM}
\end{equation}
Based on the principle of DAM~\cite{DAM}, the introduced delay ${\kappa _l}$ is set to $	{\kappa _l} = {n_{\max }} - {n_l} \ge 0,\forall l$.
Thus, the received signal in \eqref{y[n] of DAM} can be further expressed as
\begin{equation}
	\begin{split}
		y[n] =&\left( {\sum\limits_{l = 1}^L {{\bf{h}}_l^H{{\bf{F}}_{\mathrm{RF}}}{{\bf{f}}_{\mathrm{BB,}l}}} } \right)s[n - {n_{\max }}] + \\
		&\sum\limits_{l = 1}^L {\sum\limits_{l' \ne l}^L {{\bf{h}}_l^H{{\bf{F}}_{\mathrm{RF}}}{{\bf{f}}_{\mathrm{BB,}l'}}s[n - {n_{\max }} + {n_{l'}} - {n_l}]} }  + z[n].
	\end{split}
	\label{y[n] with delay compensation}
\end{equation}
It can be observed from \eqref{y[n] with delay compensation} that if the receiver is locked to the delay ${n_{\max }}$, the first term contributes to the desired signal, while the second term represents the ISI. Furthermore, the ISI can be eliminated by jointly designing $\bf{F}_{\mathrm{RF}}$ and $\bf{F}_{\mathrm{BB}}$ to satisfy
\begin{equation}
	{\bf{h}}_l^H{{\bf{F}}_{\mathrm{RF}}}{{\bf{f}}_{\mathrm{BB,}l'}} = 0,\forall l' \ne l.
	\label{ISI-ZF}
\end{equation}
Then, the received signal in \eqref{y[n] with delay compensation} reduces to
\begin{equation}
	y[n] = \left( {\sum\limits_{l = 1}^L {{\bf{h}}_l^H{{\bf{F}}_{\mathrm{RF}}}{{\bf{f}}_{\mathrm{BB,}l}}} } \right)s[n - {n_{\max }}] + z[n].
	\label{y[n] without ISI}
\end{equation}
It can be seen from \eqref{y[n] without ISI} that if the condition \eqref{ISI-ZF} is satisfied, the ISI can be eliminated perfectly and the received signal becomes a symbol sequence with the delay $n_{\max}$ and multiplied by the gains of $L$ multi-path contributions. The method in \eqref{ISI-ZF} is referred to as ISI-zero-forcing (ISI-ZF) beamforming. In this case, the signal-to-noise ratio (SNR) is
\begin{equation}
	\gamma  = \frac{1}{{{\sigma ^2}}}{\left| {\sum\limits_{l = 1}^L {{\bf{h}}_l^H{{\bf{F}}_{\mathrm{RF}}}{{\bf{f}}_{\mathrm{BB,}l}}} } \right|^2}.
	\label{SNR}
\end{equation}

Therefore, the analog and digital beamforming matrices ${\bf{F}}_{\mathrm{RF}}$ and ${\bf{F}}_{\mathrm{BB}}$ can be jointly designed to maximize the SNR in \eqref{SNR} with the ISI-ZF constraint in \eqref{ISI-ZF}. The problem can be stated as
\begin{equation}
	\begin{split}
		\mathop {\max }\limits_{{{\bf{F}}_{\mathrm{RF}}},{{\bf{F}}_{\mathrm{BB}}}} \enspace&{\left| {\sum\limits_{l = 1}^L {{\bf{h}}_l^H{{\bf{F}}_{\mathrm{RF}}}{{\bf{f}}_{\mathrm{BB,}l}}} } \right|^2}\\
		\text{s.t.}\enspace&{\bf{h}}_l^H{{\bf{F}}_{\mathrm{RF}}}{{\bf{f}}_{\mathrm{BB,}l'}} = 0,\forall l' \ne l,\\
		&\left| {{{\left( {{{\bf{F}}_{\mathrm{RF}}}} \right)}_{i,j}}} \right| = 1,\forall i,j,\\
		&\left\| {{{\bf{F}}_{\mathrm{RF}}}{{\bf{F}}_{\mathrm{BB}}}} \right\|_F^2 \le P.
	\end{split}
	\label{DAM with Hybrid}
\end{equation}
Directly solving the optimization problem \eqref{DAM with Hybrid} is difficult, since it is non-convex  and the analog and digital beamforming vectors are closely coupled. By following the similar idea of hybrid beamforming optimization as \cite{OMP}, we may first obtain the optimal solution for fully digital beamforming design, and then find the hybrid analog/digital beamforming matrices to approach the optimal fully digital beamforming as closely as possible. Specifically, by letting  ${{\bf{F}}_{\mathrm{RF}}}{{\bf{f}}_{\mathrm{BB,}l}} = {{\bf{f}}_l},\forall l$, a new beamforming problem for DAM can be formulated as
\begin{equation}
	\begin{split}
		\mathop {\max }\limits_{\left\{ {{\bf{f}_l}} \right\}_{l = 1}^L} \enspace&{\left| {\sum\limits_{l = 1}^L {{\bf{h}}_l^H{{\bf{f}}_l}} } \right|^2}\\
		\text{s.t.}\enspace&{\bf{h}}_l^H{{\bf{f}}_{l'}} = 0,\forall l' \ne l,\\
		&\sum\limits_{l = 1}^L {{{\left\| {{{\bf{f}}_l}} \right\|}^2}}  \le P.
	\end{split}
	\label{DAM with digital}
\end{equation}

Note that problem \eqref{DAM with digital} corresponds to ISI-ZF for DAM with fully digital beamforming, whose optimal solution $\{{\bf{f}}_l^{\mathrm{opt}}\}_{l=1}^L$ has been obtained in \cite{DAM}. To get an effective solution to \eqref{DAM with Hybrid}, we then need to find the hybrid beamforming design by letting ${{\bf{F}}_{\mathrm{RF}}}{{\bf{f}}_{\mathrm{BB,}l}}$ approach to ${\bf{f}}_l^{\mathrm{opt}}, \forall l$.

Let ${{\bf{H}}_l} = [{{\bf{h}}_1}, \cdots ,{{\bf{h}}_{l' - 1}},{{\bf{h}}_{l' + 1}}, \cdots {{\bf{h}}_L}],\forall l$, and the projection matrix into the space orthogonal to the columns of ${{\bf{H}}_l}$ can be expressed as: ${{\bf{Q}}_l} \buildrel \Delta \over = {{\bf{I}}_{{M_\mathrm{t}}}} - {{\bf{H}}_l}{({\bf{H}}_l^H{{\bf{H}}_l})^{ - 1}}{\bf{H}}_l^H$. According to the ISI-ZF beamforming in~\cite{DAM},  the optimal solution to the problem \eqref{DAM with digital} is
\begin{equation}
	{\bf{f}}_l^{\mathrm{opt}} = {{\sqrt P {{\bf{Q}}_l}{{\bf{h}}_l}}} \bigg/ {{\sqrt {\sum\nolimits_{l = 1}^L {{{\left\| {{{\bf{Q}}_l}{{\bf{h}}_l}} \right\|}^2}} } }},\forall l.
	\label{f_l}
\end{equation}

In order to achieve the same performance as fully digital beamforming based DAM with hybrid beamforming, ${{\bf{F}}_{\mathrm{RF}}}$ and ${{\bf{F}}_{\mathrm{BB}}}$ should be jointly designed to satisfy
\begin{equation}
	{{\bf{F}}_{\mathrm{RF}}}{{\bf{f}}_{\mathrm{BB,}l}} = {\bf{f}}_l^{\mathrm{opt}},\forall l.
	\label{hybrid design condition}
\end{equation}

 By letting ${{\bf{F}}_{\mathrm{opt}}} = [{\bf{f}}_1^{\mathrm{opt}},{\bf{f}}_2^{\mathrm{opt}}, \ldots {\bf{f}}_L^{\mathrm{opt}}]{ \in \mathbb{C}^{{M_\mathrm{t}} \times L}}$, \eqref{hybrid design condition} can be compactly expressed as: ${{\bf{F}}_{\mathrm{RF}}}{{\bf{F}}_{\mathrm{BB}}}{\rm{ = }}{{\bf{F}}_{\mathrm{opt}}}$. It can be concluded from \eqref{f_l} that the rank of ${\bf{F}}_{\mathrm{opt}}$ is $L$ when the vectors ${\bf{h}}_l,\forall l$ are linearly independent. On the other hand, the rank of ${{\bf{F}}_{\mathrm{RF}}}{{\bf{F}}_{\mathrm{BB}}}$ is no greater than ${M_{\mathrm{RF}}}$. This indicates that for \eqref{hybrid design condition} to hold, one necessary condition is ${M_{\mathrm{RF}}} \ge L$~\cite{M_RF,M_RF2}. 
Furthermore, it was revealed in \cite{M_RF2} that when ${M_{\mathrm{RF}}} \geq 2L$, ${{\bf{F}}_{\mathrm{RF}}}$ and ${{\bf{F}}_{\mathrm{BB}}}$ can be found to satisfy ${{\bf{F}}_{\mathrm{RF}}}{{\bf{F}}_{\mathrm{BB}}}{\rm{ = }}{{\bf{F}}_{\mathrm{opt}}}$ exactly, indicating that hybrid beamforming based DAM can fully realize the performance of fully digital beamforming when ${M_{\mathrm{RF}}} \ge 2L$.
Therefore, in the following, we focus on the case when ${M_{\mathrm{RF}}} < 2L$.

By following the same method as \cite{OMP}, hybrid beamforming for DAM can be designed by considering the following optimization problem:
\begin{equation}
	\begin{split}
			\mathop {\min}\limits_{{{\bf{F}}_{\mathrm{RF}}}{\rm{,}}{{\bf{F}}_{\mathrm{BB}}}} \enspace &{\left\| {{{\bf{F}}_{\mathrm{opt}}} - {{\bf{F}}_{\mathrm{RF}}}{{\bf{F}}_{\mathrm{BB}}}} \right\|_F}\\
			\text{s.t.} \enspace &\left| {{{\left( {{{\bf{F}}_{\mathrm{RF}}}} \right)}_{i,j}}} \right| = 1,\forall i,j\\
			&\left\| {{{\bf{F}}_{\mathrm{RF}}}{{\bf{F}}_{\mathrm{BB}}}} \right\|_F^2 = P.
	\end{split}
	\label{hubrid problem 1}
\end{equation}
Problem \eqref{hubrid problem 1} can be regarded as a matrix factorization problem, and it implies that ${{\bf{F}}_{\mathrm{RF}}}$ and ${{\bf{F}}_{\mathrm{BB}}}$ should be designed so that the linear combinations of the columns of ${{\bf{F}}_{\mathrm{RF}}}$ can be as close to the columns of ${{\bf{F}}_{\mathrm{opt}}}$ as possible. However, the unit modulus constraint of ${{\bf{F}}_{\mathrm{RF}}}$ makes finding the optimal solutions both analytically and algorithmically intractable~\cite{OMP}.
To find the near-optimal solution for the problem \eqref{hubrid problem 1}, the structure of mmWave massive MIMO channels can be exploited~\cite{OMP}. On the one hand, based on the expression of ${\bf{f}}_l^{\mathrm{opt}}$ in \eqref{f_l} and the channel vectors in \eqref{hl}, it can be observed that the optimal fully digital beamforming matrix ${{\bf{F}}_{\mathrm{opt}}}$ is related to the transmit array response vectors ${{\bf{a}}_\mathrm{t}}({\theta _{li}}),\forall l,i$. On the other hand, it is clear from \eqref{array response} that transmit array response vectors ${{\bf{a}}_\mathrm{t}}({\theta _{li}}),\forall l,i$ satisfy the unit modulus constraint, and the vectors are linearly independent when the AoDs ${\theta _{li}}$ are distinct, because the matrix formed by ${{\bf{a}}_\mathrm{t}}({\theta _{li}})$ with distinct ${\theta _{li}}$ will be a Vandermonde matrix with full rank.

Therefore, it is a natural idea to select $M_\mathrm{RF}$ vectors from the transmit array response vectors to construct the analog beamforming matrix ${{\bf{F}}_{\mathrm{RF}}}$. Thus, by letting ${{\bf{f}}_{\mathrm{RF},j}} \in \{{{\bf{a}}_\mathrm{t}}({\theta _{li}}),\forall l,i\}$, where ${{\bf{f}}_{\mathrm{RF},j}}$ denotes the $j$th column of ${{\bf{F}}_{\mathrm{RF}}}$ and ${{\bf{a}}_\mathrm{t}}({\theta _{li}}),\forall l,i$ correspond to the transmit array response vectors in \eqref{hl}, problem in \eqref{hubrid problem 1} can be further stated as
\begin{equation}
	\begin{split}
			\mathop {\min}\limits_{{{\bf{F}}_{\mathrm{RF}}}{\rm{,}}{{\bf{F}}_{\mathrm{BB}}}} \enspace&\left\| {{{\bf{F}}_{\mathrm{opt}}} - {{\bf{F}}_{\mathrm{RF}}}{{\bf{F}}_{\mathrm{BB}}}} \right\|_F\\
			\text{s.t.}\enspace&{{\bf{f}}_{\mathrm{RF},j}} \in {{\bf{a}}_\mathrm{t}}({\theta _{li}}),\forall l,i,j\\
			&{\left\| {{{\bf{F}}_{\mathrm{RF}}}{{\bf{F}}_{\mathrm{BB}}}} \right\|_F^2} = P.
	\end{split}
	\label{hybrid problem 2}
\end{equation}

It is well known that one efficient algorithm for problem \eqref{hybrid problem 2} is the orthogonal matching pursuit (OMP) algorithm, whose details can be found in Algorithm 1 of~\cite{OMP} and are omitted here for brevity.

Due to the unit modulus constraint for the analog beamforming matrix, the performance of hybrid beamforming may be different from the optimal fully digital beamforming when ${M_{\mathrm{RF}}} < 2L$. In particular, with the aforementioned hybrid analog/digital beamforming design, there is no guarantee that the ISI-ZF constraint \eqref{ISI-ZF} is satisfied. This implies that in order to evaluate the resulting performance of hybrid analog/digital beamforming based DAM, the residual ISI needs to be taken into account. 

To this end, the same delay components in \eqref{y[n] with delay compensation} should be grouped together~\cite{DAM}. Specifically, let ${\cal L} \buildrel \Delta \over = \left\{ {l:l = 1, \ldots ,L} \right\}$ denote the set of all multi-paths, and ${{\cal L}_l} \buildrel \Delta \over = {\cal L}\backslash l$ represents the subset that excludes the $l$th path. Furthermore, let ${\Delta _{l',l}} \buildrel \Delta \over = {n_{l'}} - {n_l}$ denotes the delay difference between $l'$ and $l$. Then, for $\forall l \ne l'$, ${\Delta _{l',l}} \in \left\{ { \pm 1, \ldots , \pm {n_{\mathrm{span}}}} \right\}$. Thus, \eqref{y[n] with delay compensation} can be equivalently expressed as
\begin{equation}
	\begin{split}
		\hspace{-1ex} y[n] = &\left( {\sum\limits_{l = 1}^L {{\bf{h}}_l^H{{\bf{F}}_{\mathrm{RF}}}{{\bf{f}}_{\mathrm{BB,}l}}} } \right)s[n - {n_{\max }}] + \\
		&\sum\limits_{l = 1}^L {\sum\limits_{l' \ne l}^L {{\bf{h}}_l^H{{\bf{F}}_{\mathrm{RF}}}{{\bf{f}}_{\mathrm{BB,}l'}}s[n - {n_{\max }} + {\Delta _{l',l}}]} }  + z[n].
	\end{split}
	\label{y[n] with delta}
\end{equation}
The terms with the same delay difference in \eqref{y[n] with delta} correspond to the same symbols, which need to be grouped. For each delay difference $i \in \left\{ { \pm 1, \ldots , \pm {n_{\mathrm{span}}}} \right\}$, the effective channel can be defined as
\begin{equation}
	{\bf{g}}_{l'}^H[i] \buildrel \Delta \over = \left\{ \begin{array}{l}
		{\bf{h}}_l^H, \text{if} \enspace \exists l \in {{\cal L}_{l'}},\enspace \text{s.t.}\enspace{n_{l'}} - {n_l} = i,\\
		{\bf{0}},\text{otherwise}.
	\end{array} \right.
\end{equation}
Therefore, \eqref{y[n] with delta} can be equivalently expressed as
\begin{equation}
	\begin{split}
			y[n] = &\left( {\sum\limits_{l = 1}^L {{\bf{h}}_l^H{{\bf{F}}_{\mathrm{RF}}}{{\bf{f}}_{\mathrm{BB,}l}}} } \right)s[n - {n_{\max }}] + \\
			&\sum\limits_{i =  - {n_{\mathrm{span}}},i \ne 0}^{{n_{\mathrm{span}}}} {\left( {\sum\limits_{l' = 1}^L {{\bf{g}}_{l'}^H[i]{{\bf{F}}_{\mathrm{RF}}}{{\bf{f}}_{\mathrm{BB,}l'}}} } \right)s[n - {n_{\max }} + i]} + \\ &z[n].
	\end{split}
\end{equation}
Then, the signal-to-interference-plus-noise ratio (SINR) of DAM with hybrid beamforming is
\begin{equation}
	{\gamma _{\mathrm{DAM\_hybrid}}} = \frac{{{{\left| {\sum\nolimits_{l = 1}^L {{\bf{h}}_l^H{{\bf{F}}_{\mathrm{RF}}}{{\bf{f}}_{\mathrm{BB,}l}}} } \right|}^2}}}{{\sum\nolimits_{i =  - {n_{\mathrm{span}}},i \ne 0}^{{n_{\mathrm{span}}}} {{{\left| {\sum\nolimits_{l' = 1}^L {{\bf{g}}_{l'}^H[i]{{\bf{F}}_{\mathrm{RF}}}{{\bf{f}}_{\mathrm{BB,}l'}}} } \right|}^2}}  + {\sigma ^2}}}.
	\nonumber
	\label{SNR of DAM with Hybrid}
\end{equation}

\section{DAM versus OFDM with Hybrid Beamforming}
\begin{figure}[t]
	\centerline{\includegraphics[width=8.5cm]{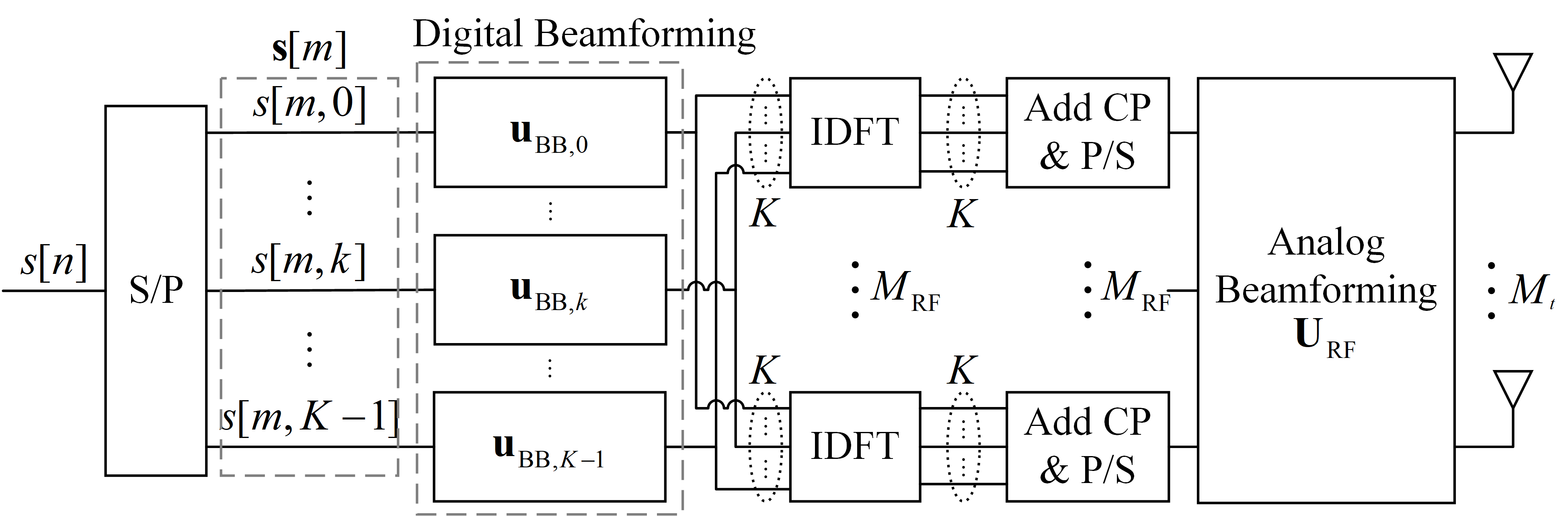}}
	\caption{Transmitter architecture for OFDM based on hybrid beamforming.}
	\label{OFDM}
\end{figure}

In this section, we compare the hybrid beamforming based DAM with OFDM. The transmitter architecture of OFDM based on hybrid analog/digital beamforming is illustrated in Fig.\ref{OFDM}, and the transmitted signal can be expressed as
\begin{equation}
	{{\bf{\bar x}}_\mathrm{t}}[m,n] = \frac{1}{{\sqrt K }}{{\bf{U}}_{{\mathrm{RF}}}}\sum\limits_{k = 0}^{K - 1} {{{\bf{u}}_{{\mathrm{BB}},k}}s[m,k]{e^{j\frac{{2\pi }}{K}kn}}},
	\label{x[m,n]}
\end{equation}
where $K$ denotes the number of subcarriers used in OFDM, ${{\bf{U}}_{\mathrm{RF}}}{ \in \mathbb{C}^{{M_\mathrm{t}} \times {M_{\mathrm{RF}}}}}$ denotes the analog beamforming matrix, ${{\bf{u}}_{{\mathrm{BB}},k}}$ denotes the digital baseband beamforming vector for subcarrier $k$, $s[m,k]$ denotes the information symbol carried by the $k$th subcarrier in the $m$th OFDM symbol, and $n =  - {N_\mathrm{CP}},- {N_\mathrm{CP}}+1, \cdots ,K - 1$, where $N_\mathrm{CP}$ denotes the length of CP.

Through the channel in \eqref{Channel} , the received signal can be expressed as
\begin{equation}
	\begin{split}
		&{{{{\bar y}}}_\mathrm{t}}[m,n] \\&= \sum\limits_{l = 1}^L {{\bf{h}}_l^H} \frac{1}{{\sqrt K }}{{\bf{U}}_{{\rm{RF}}}}\sum\limits_{k = 0}^{K - 1} {{{\bf{u}}_{{\rm{BB}},k}}s[m,k]{e^{j\frac{{2\pi }}{K}k(n - {n_l})}}} \\&+ z[m,n]\\
		&= \frac{1}{{\sqrt K }}\sum\limits_{k = 0}^{K - 1} {\left( {\sum\limits_{l = 1}^L {{\bf{h}}_l^H{e^{ - j\frac{{2\pi }}{K}k{n_l}}}} } \right){{\bf{U}}_{{\rm{RF}}}}{{\bf{u}}_{{\rm{BB}},k}}s[m,k]{e^{j\frac{{2\pi }}{K}kn}}} \\&+ z[m,n],
	\end{split}
	\label{y[m,n]}
\end{equation}
where $z[m,n] \sim {\cal C}{\cal N}(0,{\sigma ^2})$ is the AWGN.
With discrete Fourier transform (DFT), the frequency-domain channel of the $k$th subcarrier can be obtained from \eqref{Channel} as follows:
\begin{equation}
	{{\bf{h}}^H}[k] = \frac{1}{{\sqrt K }}\sum\limits_{l = 1}^L {{\bf{h}}_l^H{e^{ - j\frac{{2\pi }}{K}k{n_l}}}}.
	\label{h[k]}
\end{equation}
By substituting \eqref{h[k]} into \eqref{y[m,n]}, we can obtain
\begin{equation}
	\begin{split}
	\hspace{-1.59ex} {\bar y_\mathrm{t}}[m,n] &= \frac{1}{{\sqrt K }}\sum\limits_{k = 0}^{K - 1} {\sqrt K {{\bf{h}}^H}[k]{{\bf{U}}_{{\rm{RF}}}}{{\bf{u}}_{{\rm{BB}},k}}s[m,k]{e^{j\frac{{2\pi }}{K}kn}}} \\&+z[m,n].
	\end{split}
\end{equation}
After removing the CP and performing DFT to (27), we can obtain
\begin{equation}
	y[m,k] =\sqrt{K}{\bf{h}}{[k]^H}{{\bf{U}}_{\mathrm{RF}}}{{\bf{u}}_{\mathrm{BB},k}}s[m,k] + z[m,k].
\end{equation}
Therefore, the received SNR is given by ${\gamma _{\mathrm{OFDM\_hybrid}}} = {{{{\left| {{\bf{h}}^H{{[k]}}{{\bf{U}}_{\mathrm{RF}}}{{\bf{u}}_{\mathrm{BB},k}}} \right|}^2}}} / \left(\sigma^2/K\right)$.

For maximizing the SNR, ${\bf{U}}_{\mathrm{RF}}$ and ${\bf{U}}_{\mathrm{BB}}$ can be jointly designed to approach the per-subcarrier based fully digital MRT beamforming
\begin{equation}
	{\bf{u}}_k^{\mathrm{opt}} = \sqrt {{p_k}} {{{\bf{h}}[k]}}/{{\left\| {{\bf{h}}[k]} \right\|}},
	\label{U}
\end{equation}
where $p_k$ denotes the transmit power of the $k$th subcarrier, which can be obtained by the water-filling (WF) power allocation.
Let the optimal fully digital beamforming matrix ${{\bf{U}}_{\mathrm{opt}}} = [{\bf{u}}_1^{\mathrm{opt}},{\bf{u}}_2^{\mathrm{opt}}, \ldots {\bf{u}}_K^{\mathrm{opt}}]{ \in \mathbb{C}^{{M_\mathrm{t}} \times K}}$ and ${\bf{H}} = [{\bf{h}}[0],{\bf{h}}[1], \ldots {\bf{h}}[K - 1]]{ \in ^{{M_\mathrm{t}} \times K}}$. Since DFT is a linear transformation, it can be shown that the rank of ${\bf{H}}$ is $L$ when the vectors ${\bf{h}}_l,\forall l$ are linearly independent. Thus, the rank of ${{\bf{U}}_{\mathrm{opt}}}$ can be concluded to be $L$ with \eqref{U}. Therefore, similar to DAM, for hybrid beamforming based OFDM to achieve the same performance as fully digital beamforming, it is required to satisfy ${M_{\mathrm{RF}}} \ge L$. Besides, when  $ {M_{\mathrm{RF}}} < 2L$, the problem of the hybrid beamforming design for OFDM can be formulated as
\begin{equation}
	\begin{split}
		\mathop {\min}\limits_{{{\bf{F}}_{\mathrm{RF}}}{\rm{,}}{{\bf{F}}_{\mathrm{BB}}}} &\left\| {{{\bf{U}}_{\mathrm{opt}}} - {{\bf{U}}_{\mathrm{RF}}}{{\bf{U}}_{\mathrm{BB}}}} \right\|_F\\
		\text{s.t.}\enspace &\left| {{{\left( {{{\bf{U}}_{\mathrm{RF}}}} \right)}_{i,j}}} \right| = 1,\forall i,j\\
		&{\left\| {{{\bf{U}}_{\mathrm{RF}}}{{\bf{U}}_{\mathrm{BB}}}} \right\|_F^2} = P.
	\end{split}
\end{equation}
The problem above is similar to the hybrid beamforming design for DAM, which can also be solved with the OMP algorithm in \cite{OMP}.
 
Let $T_c$ denote the channel coherence time, during which the channel remains approximately unchanged. Then the number of single-carrier symbols within each channel coherence block is given by ${n_c} \approx {{{T_c}} \mathord{\left/{\vphantom {{{T_c}} {{T_s}}}} \right.\kern-\nulldelimiterspace} {{T_s}}}$, where ${T_s} = {1 \mathord{\left/
{\vphantom {1 B}} \right. \kern-\nulldelimiterspace} B}$ denotes the symbol duration with available bandwidth $B$. Let ${\tilde n_{\max }}$ denotes the upper bound of the maximum delay over all channel coherence blocks, i.e., ${\tilde n_{\max }} \ge {n_{\max }}$. To avoid ISI across different coherence blocks, it is necessary to insert the guard interval for each coherence block. It can be observed from \eqref{y[n] with delay compensation} that for DAM, each coherent block requires a guard interval of length $2{\tilde n_{\max }}$, which corresponds to a guard interval overhead of ${{2{{\tilde n}_{\max }}} \mathord{\left/
{\vphantom {{2{{\tilde n}_{\max }}} {{n_c}}}} \right.
\kern-\nulldelimiterspace} {{n_c}}}$. Therefore, the effective spectral efficiency of DAM in (bps/Hz) can be expressed as
\begin{equation}
	{R_{\mathrm{DAM}}} = \frac{{{n_c} - 2{{\tilde n}_{\max }}}}{{{n_c}}}{\log _2}(1 + {\gamma _\alpha }),
	\label{RDAM}
\end{equation}
where $\alpha  \in \left\{ {\mathrm{DAM\_digital, DAM\_hybrid}} \right\}$.

For OFDM with $K$ subcarriers, in order to avoid ISI across different OFDM symbols, a CP of length ${\tilde n_{\max }}$ needs to be inserted for each OFDM symbol. The number of OFDM symbols within each channel coherence block is ${n_{\mathrm{OFDM}}} = {n_c}/(K + {\tilde n_{\max }})$, and the guard interval overhead can be calculated as ${{{n_{\mathrm{OFDM}}}{{\tilde n}_{\max }}} \mathord{\left/	{\vphantom {{{n_{\mathrm{OFDM}}}{{\tilde n}_{\max }}} {{n_c}}}} \right.		\kern-\nulldelimiterspace} {{n_c}}}$. Thus, the effective spectral efficiency of OFDM with MRT beamforming can be expressed as
\begin{equation}
	{R_{\mathrm{OFDM}}} = \frac{{{n_c} - {n_{\mathrm{OFDM}}}{{\tilde n}_{\max }}}}{{{n_c}}}\frac{1}{K}\sum\limits_{k = 1}^K {{{\log }_2}(1 + {\gamma _\beta })},
	\label{ROFDM} 
\end{equation}
where $\beta  \in \left\{ {\mathrm{OFDM\_digital, OFDM\_hybrid}} \right\}$. 
It can be inferred that ${{2{{\tilde n}_{\max }}} \mathord{\left/{\vphantom {{2{{\tilde n}_{\max }}} {{n_c}}}} \right.
\kern-\nulldelimiterspace} {{n_c}}}{{ \ll {n_{\mathrm{OFDM}}}{{\tilde n}_{\max }}} \mathord{\left/
{\vphantom {{ \ll {n_{\mathrm{OFDM}}}{{\tilde n}_{\max }}} {{n_c}}}} \right.
\kern-\nulldelimiterspace} {{n_c}}}$ when ${n_{\mathrm{OFDM}}} \gg 1$. This indicates that DAM can reduce the guard interval overhead significantly, compared with OFDM.

For analyzing the PAPR, let the transmit signal on each antenna be ${x_{{m_\mathrm{t}}}}[n]$, and its PAPR can be expressed as: ${{\max ({{\left| {{x_{{m_\mathrm{t}}}}[n]} \right|}^2})} \mathord{\left/
{\vphantom {{\max ({{\left| {{x_{{m_\mathrm{t}}}}[n]} \right|}^2})} {E({{\left| {{x_{{m_\mathrm{t}}}}[n]} \right|}^2})}}} \right.
\kern-\nulldelimiterspace} {\mathbb{E}[{{\left| {{x_{{m_\mathrm{t}}}}[n]} \right|}^2}]}}$. Thus, the overall PAPR is obtained by taking the maximum PAPR across all transmit antennas:
\begin{equation}
	\mathrm{PAPR} = \mathop {\max }\limits_{1 \le {m_\mathrm{t}} \le {M_\mathrm{t}}} \left( {\frac{{\mathop {\max }\limits_{1 \le n \le S} \left( {{{\left| {{x_{{m_\mathrm{t}}}}[n]} \right|}^2}} \right)}}{\mathbb{E}{\left[ {{{\left| {{x_{{m_\mathrm{t}}}}[n]} \right|}^2}} \right]}}} \right),
	\label{PAPR}
\end{equation}
where $S$ denotes the number of transmitted symbols over each block. It can be seen from \eqref{x[m,n]} that the transmitted signal of OFDM is a superposition of $K$ orthogonal subcarrier signals, and if they happen to be exactly in-phase, a large instantaneous power peak will be generated. As for DAM in \eqref{x[n]}, its PAPR is mainly affected by the number of multi-paths $L$, which is small due to the multi-path sparsity of mmWave communication. Therefore, DAM is expected to achieve a lower PAPR than OFDM.

\section{Simulation Results}

\begin{table}[!t]
	\renewcommand{\arraystretch}{1.3}
	\caption{Parameter settings}
	\centering
	\label{Parameter}
	\resizebox{\columnwidth}{!}{
		\begin{tabular}{l l}
			\hline\hline \\[-4mm]
			\multicolumn{1}{l}{Parameter} & \multicolumn{1}{l}{value} \\[0.5ex] \hline
			Number of transmit antennas & ${M_\mathrm{t}} = 256$ \\
			Number of RF chains & ${M_\mathrm{RF}} = 5$ \\
			Carrier frequency & $f = 28\mathrm{GHz}$ \\
			Total bandwidth & $B = 128\mathrm{MHz}$ \\
			Noise power spectral density & ${N_0} =  - 174\mathrm{dBm/Hz}$\\
			Inter-element spacing of the ULA & $d = {\lambda  \mathord{\left/{\vphantom {\lambda  2}} \right.\kern-\nulldelimiterspace} 2}$\\
			Channel coherence time & ${T_c} = 1\mathrm{ms}$\\
			Number of temporal-resolvable multi-paths & $L = 5$\\
			Delay & ${n_l} \sim \mathrm{U}[0,{\tau _{\max }}],{\tau _{\max }} = 312.5\mathrm{ns}$\\
			Number of sub-paths & ${\mu _l} \sim \mathrm{U}[0,{\mu _{\max }}],{\mu _{\max }} = 3$\\
			AoDs & ${\theta _{li}} \sim \mathrm{U}[ - 60^\circ ,60^\circ ],\forall l,i$\\
			Number of sub-carriers used in OFDM & $K = 512$\\
			\hline\hline
		\end{tabular}
	}
\end{table}
Unless otherwise stated, the parameter settings of the simulation results are summarized in Table \ref{Parameter}. The upper bound of the maximum delay over all channel coherence blocks is ${\tilde n_{\mathrm{max}}} = {\tau _{\max }}B = 40$. The total number of single-carrier symbols and the number of OFDM symbols within each channel coherence time is ${n_c} = B{T_c} = 1.28 \times {10^5}$, and ${n_{\mathrm{OFDM}}} = \frac{{{T_c}}}{{(K + {{\tilde n}_{\mathrm{max}}}){T_s}}} \approx 231$, respectively. Furthermore, the complex-valued gains ${\alpha _l},\forall l$ in \eqref{hl} are generated based on the model developed in~\cite{Sparsity-1}.

\begin{figure}[t]
	\centerline{\includegraphics[width=5.8cm]{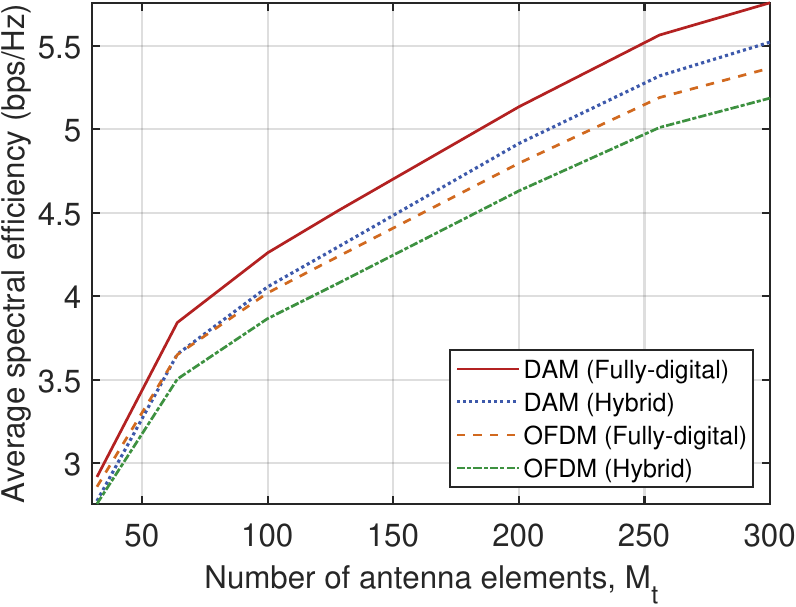}}
	\caption{Average spectral efficiency comparison for DAM and OFDM based on fully digital and hybrid beamforming.}
	\label{Spec}
\end{figure}

\begin{figure}[t]
	\centerline{\includegraphics[width=5.8cm]{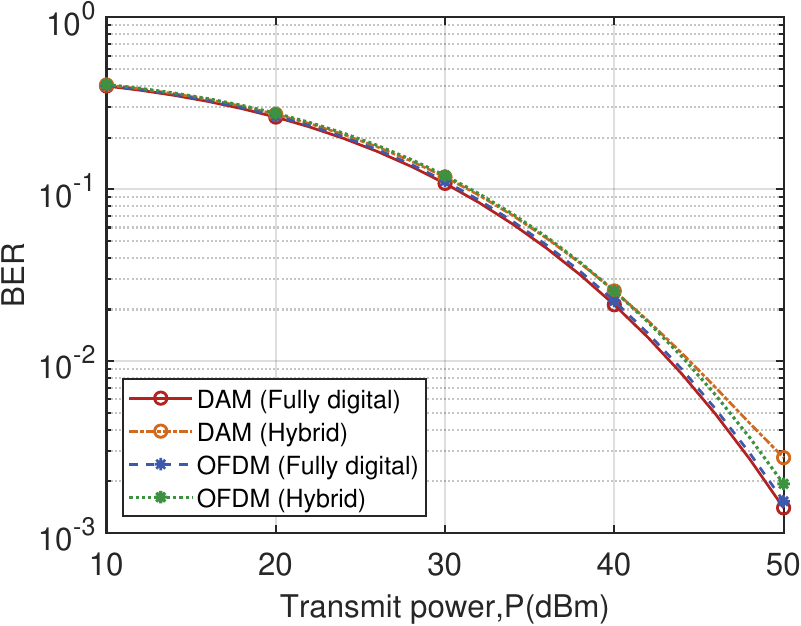}}
	\caption{BER comparison for DAM and OFDM based on fully digital and hybrid beamforming.}
	\label{BER}
\end{figure}

\begin{figure}[t]
	\centerline{\includegraphics[width=5.8cm]{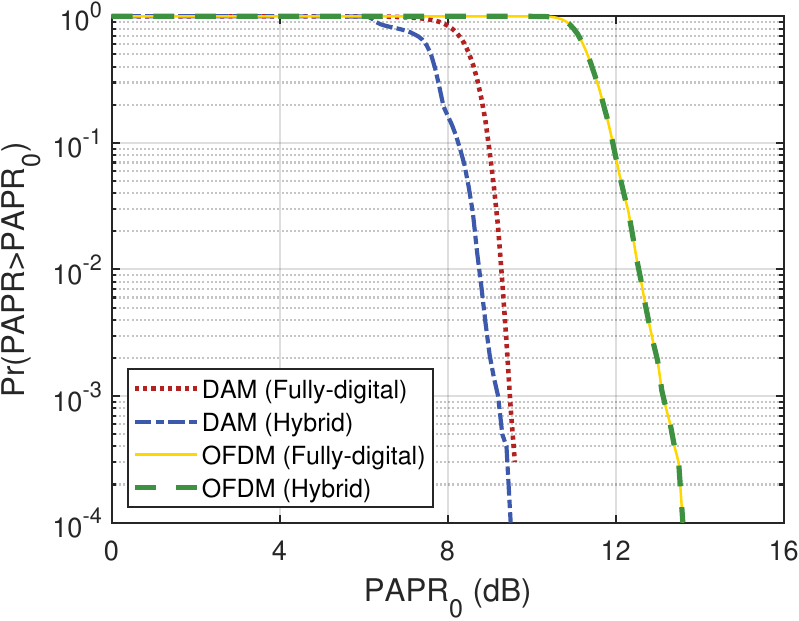}}
	\caption{PAPR comparison for DAM and OFDM based on fully digital and hybrid beamforming.}
	\label{PAPR}
\end{figure}

The effective spectral efficiency of DAM and OFDM based on different beamforming methods can be obtained with \eqref{RDAM} and \eqref{ROFDM}. For transmit power of $P = 30\mathrm{dBm}$, Fig. \ref{Spec} shows the average spectral efficiency versus the number of antennas ${M_\mathrm{t}}$ for DAM and OFDM based on fully digital and hybrid beamforming. It can be observed that for both DAM and OFDM, the average spectral efficiency achieved using hybrid beamforming is slightly lower than that by using fully digital beamforming. This is due to the unit modulus constraint imposed on the analog beamforming matrix in the hybrid scheme. It can also be observed from Fig. \ref{Spec} that for both fully digital and hybrid beamforming, DAM outperforms OFDM in terms of average spectral efficiency, since the guard interval overhead of DAM, which is ${{2{{\tilde n}_{\max }}} \mathord{\left/
{\vphantom {{2{{\tilde n}_{\max }}} {{n_c}}}} \right.\kern-\nulldelimiterspace} {{n_c}}} = \frac{{80}}{{1.28 \times {{10}^5}}} = 0.0625\% $, is much lower than that of OFDM, which is ${{{n_{\mathrm{OFDM}}}{{\tilde n}_{\max }}} \mathord{\left/
{\vphantom {{{n_{\mathrm{OFDM}}}{{\tilde n}_{\max }}} {{n_c}}}} \right.
\kern-\nulldelimiterspace} {{n_c}}} = \frac{{231 \times 40}}{{1.28 \times {{10}^5}}} = 7.22\% $.

Next, we compare the BER based on the Monte Caro simulations. Specifically, we simulate the transmission process of random sequences using 128-QAM over the randomly generated channels $10^3$ times, and the BER is obtained by taking the average over all transmissions. It is observed from Fig. \ref{BER} that the BER performance of DAM based on hybrid beamforming is close to that based on fully digital beamforming. This validates the effectiveness of hybrid beamforming for DAM, indicating its practical application in mmWave/THz massive MIMO communication scenarios. Additionally, it can be observed that with fully digital beamforming, the BER performance of DAM is comparable to that of OFDM. 

Fig. \ref{PAPR} plots the complementary cumulative distribution function (CCDF) of PAPR for DAM and OFDM based on fully digital and hybrid beamforming using 128QAM. It can be observed that the PAPR of DAM is significantly lower than that of OFDM for both fully digital and hybrid beamforming. Furthermore, it can be observed that the PAPR of OFDM based on hybrid beamforming is almost the same as that based on fully digital beamforming, while for DAM, the use of hybrid beamforming can further reduce the PAPR.

\section{Conclusion}
This paper proposed the hybrid beamforming-based DAM for spatially sparse communications. It not only leverages the high spatial resolution of large antenna arrays and the multi-path sparsity of mmWave/THz channels to achieve ISI-free single-carrier communication, but also save the number of RF chains, thus reducing the hardware cost and power consumption compared to DAM based on fully digital beamforming. Simulation results validate the effectiveness of hybrid beamforming for DAM, indicating its practical application in large-scale MIMO communications. Furthermore, it demonstrates the superiority of DAM over OFDM in terms of effective spectral efficiency and PAPR.

\end{document}